# Unexpected Thermal Conductivity Enhancement in Pillared Graphene Nanoribbon with Isotopic Resonance

Dengke Ma[1,2], Xiao Wan[1,2], and Nuo Yang[1,2,*]

[1]State Key Laboratory of Coal Combustion, Huazhong University of Science and Technology (HUST), Wuhan 430074, P. R. China

[2]Nano Interface Center for Energy (NICE), School of Energy and Power Engineering, Huazhong University of Science and Technology (HUST), Wuhan 430074, P. R. China

Electronic mail: N.Y. (nuo@hust.edu.cn)

**ABSTRACT**

Thermal transport in nanoribbon based nanostructures is critical to advancing its applications. Wave effects of phonons can give rise to controllability of heat conduction in nanostructures beyond that by particle scattering. In this paper, by introducing structural resonance, we systematically studied the thermal conductivity of graphene nanoribbon based phononic metamaterials (GNPM) through non-equilibrium molecular dynamical simulation. Interestingly, it is found that the thermal conductivity of GNPM is counter-intuitively enhanced by isotope doping, which is strong contrast to the common notion that isotope doping reduces thermal conductivity. Further mode-analysis and atomic green function calculation reveal that the unexpected increasing in thermal conductivity originates from the breaking of the resonant hybridization wave effect between the resonant modes and the propagating modes induced by isotope doping. Besides, factors including the system width and leg length can also efficiently tune the thermal conductivity of GNPM. This abnormal mechanism provides a new dimension to manipulate phonon transport in nanoribbon based nanostructures through wave effect.

## Introduction

Nanoribbons, which are patterned as thin strips of two-dimensional materials, have attracted great attention due to their novel properties that differ from their two-dimensional counterparts.[1-3] Graphene nanoribbon (GNR) can open the bandgap in semimetal graphene and give it the potential to be applied in transistors, photovoltaics, and valleytronics.[4] It has been demonstrated that the electric[1], magnetic[5] and spin-related[6] properties of GNR can be modulated by engineering the ribbon width and the edge configuration. The thermal properties of GNR are also very important both for improving the performance and reliability of devices and fundamental understanding of the physics in low-dimensional system.[7]

The thermal conductivity of pristine GNR with different length and width were studied by molecular dynamical simulation[8-10] and experiment[11]. To manipulate the thermal properties of GNR, different strategies have been put forward. The most commonly exercised approaches are based on the particle nature of phonon. It is found that edge roughness,[8,12] porous,[13,14] folding,[15] hydrogen termination,[8] random defect and doping[16-18] would effectively tune the thermal conductivity of GNR. By utilizing edge and folding effect, thermal diode[19] and adjustable thermal resistor[20] can be realized in asymmetric and folded GNR, respectively.

Another line to manipulate thermal properties of GNR is utilizing the wave nature of phonon[21], which can be realized in two distinct ways. The first is phononic crystal which is based on Bragg-scattering. By introducing periodic porous[22-24] or isotope[25,26] in GNR, wave interferences occur and provide a unique frequency band structure with the possibility of band gaps[27]. This would hugely reduce the thermal conductivity of GNR.[22,23,25,26] The second is phononic metamaterial (PM) which is based on local resonant hybridization.[28] The local resonant hybridization has the advantage to block phonon transport, and is not expected to scatter electrons.[28,29] What is more, the occurring of resonant hybridization do not need the structure to be periodic,[30] which simplifies the synthesis of PM. Previous researches about local resonant hybridization are focused on silicon based three dimensional nanostructures[28-31], to our best knowledge, no study has been performed on nanoribbons.

In this paper, we perform non-equilibrium molecular dynamical (NEMD) simulation to systematically study the thermal conductivity of graphene nanoribbon phononic metamaterials (GNPM). (as shown in Fig. 1) The dependence of thermal conductivity on both width ($W$) and leg length ($H_L$) are studied. Moreover, isotope doping effect on thermal conductivity by varying the atomic mass of legs is also performed. Vibrational eigen-modes analysis and atomistic Green's function (AGF) calculations are carried to reveal the underlying physical mechanism.

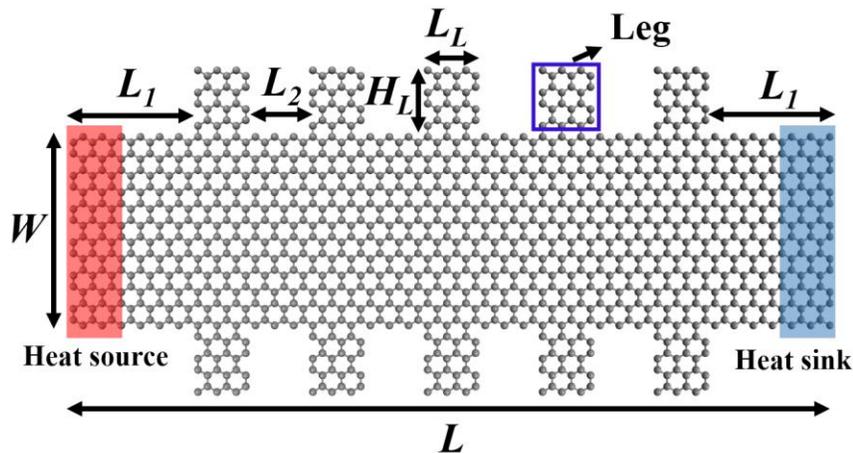

Fig.1 Schematic picture of the GNPM which is made up of GNR with legs on two sides. The fixed (free) boundary conditions are applied in longitudinal (transverse) direction. The total length ($L$) is fixed as 10 nm (80 layers). $L_1$, $L_2$ and $L_L$ are fixed as 1.62 nm, 0.87 nm and 0.62 nm, respectively.

## Structure and Methods

GNPM is made up of GNR with legs on two sides. (shown in Fig. 1) The legs serve as resonators.[28] The GNPM has been successfully fabricated of atomically precise on experiment.[6] The lattice constant (a) and thickness (d) of GNPM are 0.1438 nm and 0.334 nm, respectively. Throughout this paper, the total length ($L$), $L_1$, $L_2$ and $L_L$ are fixed as 10 nm (80 layers), 1.62 nm, 0.87 nm and 0.62 nm, respectively.

NEMD simulations in this paper are performed using LAMMPS package[32] with the optimized Tersoff potential[33], which has successfully reproduced the thermal properties of graphene[16,34,35]. The detailed parameters of optimized Tersoff potential are shown in Table 1. The fixed (free) boundary conditions are applied in longitudinal (transverse) direction. In order to establish a temperature gradient along the longitudinal direction, the system is coupled with Langevin thermostats [28] at the 3rd to 6th and (N-6)th to (N-3)th layers with 310 K and 290 K, respectively. And atoms at the boundaries (the 1st to 2nd and (N−1) th to N th layers) are fixed.

The equations of motions are integrated by velocity Verlet method with a time step of 0.5 fs. Initially, the system is relaxed in the isothermal−isobaric (NPT) ensemble at 300 K and 0 bar for 1 ns, followed by relaxation under a microcanonical ensemble (NVE) for 2.5 ns. After that, a time average of the temperature and heat current is performed for 10 ns. The results presented here are averaged over 6 independent simulations with different initial conditions, and the error bar is obtained from the standard deviation of different runs.

The temperature gradient is obtained by linear fitting to the local temperature, excluding the temperature jumps at the two ends. The thermal conductivity is calculated based on the Fourier's Law,

$$\kappa = -\frac{J}{W \cdot d \cdot \nabla T} \quad (1)$$

where $J$ denotes the heat current transported from the hot bath to cold bath, $W$ is the width, $d$ is thickness, and $\nabla T$ is the temperature gradient along the longitudinal direction. It should be noted that when calculating the cross section area of GNPM, we use the effective width ($W$). The width of the leg part does not include. The temperature $T_{MD}$ is calculated from the kinetic energy of atoms ($3Nk_B T_{MD}/2 = \sum_i m_i v_i^2/2$).

## Results and Discussion

We first calculate the $\kappa_{GNPM}$ with different $W$, and compare them with the thermal conductivity of corresponding GNR ($\kappa_{GNR}$). Here, the $H_L$ is fixed as 0.86 nm. Fig. 2 (a) shows the relative thermal conductivity ($\kappa_{GNPM}/\kappa_{GNR}$). When $W$ increases from 0.86 nm to 4.32 nm, the relative thermal conductivity (black dot) increases from 0.27 to 0.43. This means the introduction of legs (structural resonance) hugely reduce the $\kappa_{GNR}$.

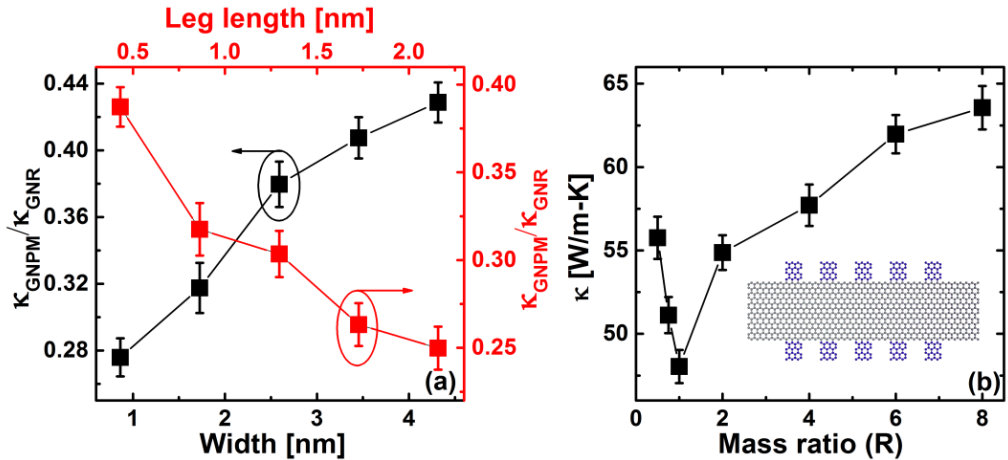

Fig.2 (a) Thermal conductivity of GNPM versus the width ($W$) and leg length ($L_L$). (b) Thermal conductivity of isotope doped GNPM with different mass ratio (R). Here, R is defined as R=M/12, where M is the atomic mass of the isotope of C in the leg (blue atoms). The mass of atoms in the ribbon part (gray atoms) is kept as 12. The inset is the schematic picture of isotope doped GNPM.

To understand the underlying physical mechanism and explicitly show the leg effect, we carry out a vibrational eigen-mode analysis of phonons in GNR and GNPM. As shown in the phonon dispersion relations of GNR (Fig.3(a)) and GNPM (Fig.3(b)), there are many flat bands in GNPM which is the signature of local resonance[28,30,36]. These resonant modes hybridize with the propagating modes of the ribbon part, which reduces the group velocity and hinder the transport of the propagating modes.[28,29] As shown in Fig.3(c), the group velocity of GNPM (red dot) are quite lower than that of GNR (blue dot) in the whole frequency range.

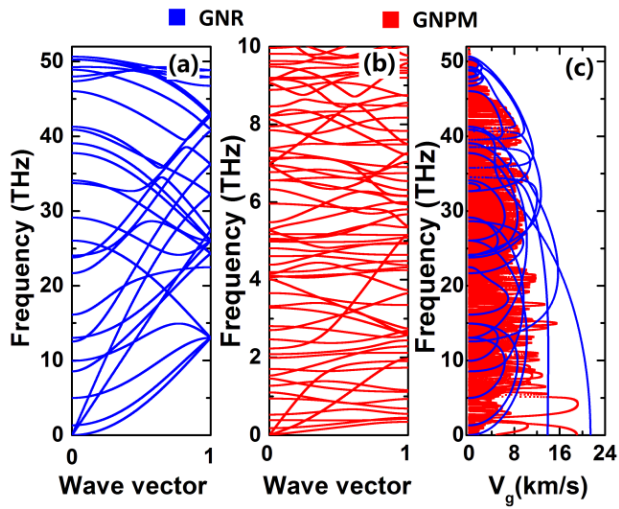

Fig.3 The phonon dispersion relationship of (a) GNR and (b) GNPM. (c) The group velocity of GNR (blue dot) and GNPM (red dot).

Since the reduction of thermal conductivity is due to the resonant hybridization wave effect, the increases of relative thermal conductivity with the increasing of $W$ is easy to be understand. As the size of the legs keep the same, the resonant modes do not change. While the number of propagating modes increases as $W$ increases. The ratio of the propagating modes that can hybridize with the resonant modes decreases, which leads to the increasing of relative thermal conductivity.

We also calculated the $\kappa_{GNPM}$ with different $H_L$, while $W$ is fixed as 1.73 nm. As shown in Fig. 2(a) (red dots), when $H_L$ increases from 0.43 nm to 2.16 nm, the $\kappa_{GNPM}/\kappa_{GNR}$ decreases from 0.39 to 0.25. As here the $\kappa_{GNR}$ does not change, it means the $\kappa_{GNPM}$ decreases with $H_L$ increases. This tendency is consistent with observations in silicon based PM.[37,38] One reason is that increasing $H_L$ increases the number of resonant modes. Another reason is that larger $H_L$ will induce resonant modes with lower frequency, and form more hybridization in low frequency.[29] Generally, the low frequency modes contribute more to the thermal conductivity.

Besides the dependence of $\kappa_{GNPM}$ on system size ($W$ and $H_L$), the isotope doping effect is also investigated. The isotope doping is an effective way to block phonon transport and reduce thermal conductivity.[39-41] There are only 15 known isotopes of C, whose atomic mass changes only from 8 to 22. Artificial C isotope atoms here are used to explore the mass influence on the thermal transport, and can be looked as other atoms,[42,43] such as $^{56}$Fe (as illustrated in Ref. 41). When there are other kind atoms, the system is more complicated. That is, the mass is not the only factor involved. The bond strength and lattice relaxations must play a role, which is not studied in this letter. The inset of Fig. 2(b) shows the structure of isotope doped GNPM. Here, we change the mass of atoms in the legs (blue atoms), and keep the mass of atoms in the ribbon part unchanged (gray atoms). The bottom-up approach is one of the best ways to fabricate the isotope doped GNPM. For example, the isotope doped GNPM can be obtained by fusing segments made from two different molecular building blocks with different atomic mass (e.g. 12 and 22).[44] Moreover, if using different kind of atoms, e.g. Fe, to instead of isotope doped C, the GNPM structure can be fabricated by the multistep method, which has been widely used in fabricating branched nanowire heterostructures.[45]

We define a parameter, mass ratio (R), as R=M/12, where M is the atomic mass of the isotope of C in the leg. The $W$ and $L_L$ are fixed as 1.73 nm and 0.86 nm, respectively. As shown in Fig. 2 (b), the minimum $\kappa_{GNPM}$ is observed when R equals to 1, which means there is no isotope doping. When R decreases from 1 to 0.5, the $\kappa_{GNPM}$ increases from 48.0 Wm$^{-1}$K$^{-1}$ to 55.8 Wm$^{-1}$K$^{-1}$. And when R increases from 1 to 8, the $\kappa_{GNPM}$ increases from 48.0 Wm$^{-1}$K$^{-1}$ to 63.6 Wm$^{-1}$K$^{-1}$. This means the isotope doping with lighter or heavier atoms both increase the $\kappa_{GNPM}$ surprisingly.

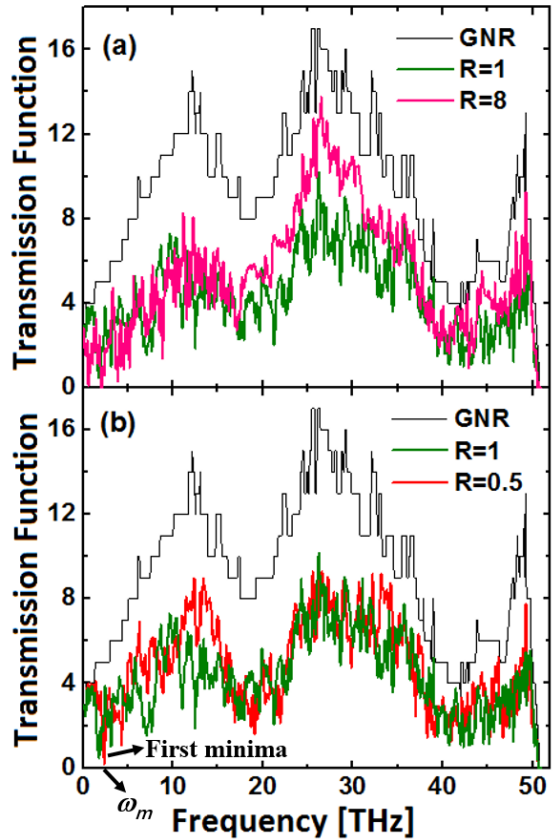

Fig.4 The phonon transmission coefficient of GNR (black line), GNPM (R=0.5) (red line), GNPM (R=1) (green line) and GNPM (R=8) (pink line).

It is well-known that the isotope doping generally induces phonon scattering and reduces thermal conductivity.[39,40] And previous studies on GNR all observed a reduction of thermal conductivity by isotope doping.[18,25,26] To understand the unexpectedly increase of $\kappa_{GNPM}$ by isotope doping, we calculate the phonon transmission for GNR and GNPM (R=0.5, 1, 8) by using the AGF method. As shown in Fig. 4 (a), the transmission function of GNR (black line) is larger than that of GNPM with no isotope doping (green line) in a wide frequency range, which also serves as an evidence for the reduction of thermal conductivity by resonant hybridization. What is more importantly, the transmission function for $R$ equals to 8 (pink line) is larger than that of pristine GNPM (R=1) in a wide high frequency range. (shown in Fig. 4(a)) And the transmission function for $R$ equals to 0.5 (red line) is larger than that of pristine GNPM (R=1) in a wide low frequency range. (shown in Fig. 4(b)) This means heavy mass isotope doping will facilitate the transport of high frequency phonons, and the light mass isotope doping will facilitate the transport of low frequency phonons. Thus the isotope doping enhances the $\kappa_{GNPM}$.

The unexpected facilitating of phonon transport by isotope doping is due to the special physical mechanism here. The local resonant hybridization wave effect mainly govern the phonon transport in GNPM which attributes to the interaction of resonant phonons in the legs with the propagate phonons in the ribbon part. The interaction (hybridization) occurs when two modes with the same polarization located on different objects are coupled at the same frequency.[36] Phonon frequency has an inverse relationship with the atomic mass.[46] Heavy mass isotope doping (R=8) will turn high frequency resonant modes to low frequency, which result in more resonant modes in low frequency. Thus, the transmission function for *R* equals to 8 (pink line) are smaller than that for pristine GNPM (R=1) in some low frequency ranges, and larger than that for pristine GNPM (R=1) in some high frequency ranges. Light mass isotope doping (R=0.5) will turn low frequency resonant modes to high frequency, which result in more resonant modes in high frequency. Thus, the transmission function for *R* equals to 0.5 (red line) are larger than that for pristine GNPM (R=1) in some low frequency ranges, and smaller than that for pristine GNPM (R=1) in some high frequency ranges. When atoms in the leg has the same mass with the atoms in the ribbon (R=1), the resonant modes will more likely to hybridize with the propagating modes of the ribbon part. After introducing isotope doping and change the mass of the legs, it breaks the original perfect hybridization and increases the $\kappa_{GNPM}$.

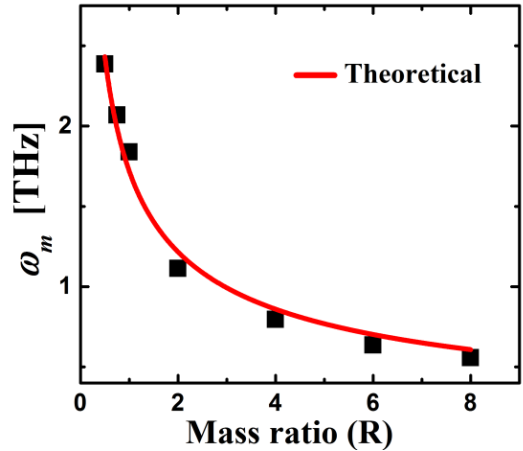

Fig.5 The frequency ($\omega_m$) of the first transmission function minima versus mass ratio.

To quantitatively show the isotope doping effect on the resonant hybridization, we focus on the shifting of the frequency ($\omega_m$) of the first transmission function minima (shown in Fig.4) by changing the mass of isotope atoms. As shown in Fig. 5, the $\omega_m$ decreases with the increasing of mass ratio (black dot), which means the first transmission function minima shift to low frequency gradually with the doped atomic mass increases. The relationship between the phonon frequency of the leg (resonant part) and the atomic mass can be approximated as $\omega \sim m^{-0.5}$.[46] We use the formula $\omega_m = aR^{-0.5}$ to fit the simulation dates, as shown in Fig. 5, the theoretical formula (red line) matches well with the simulation dates (black dot). This result further validates the behind mechanism that the isotope doping changes the frequency of resonant modes and breaks the original perfect hybridization and increases the $\kappa_{GNPM}$.

## Conclusion

In summary, we have studied the thermal conductivity of GNPM by using the NEMD simulation. It is found that the thermal conductivity of GNPM is significantly reduced as compared with the corresponding GNR (lowest relative thermal conductivity $\kappa_{GNPM}/\kappa_{GNR}$ =0.25). After comparatively studying the phonon eigen-modes in GNPM and GNR by lattice dynamics, it is demonstrated that the phonon local resonant hybridization is responsible for the reduction. It is also found that the thermal conductivity of GNPM decreases with increasing leg length ($L_L$) and decreasing width ($W$). What's more interesting, it is found that light and heavy mass isotope doping of the leg atoms can both enhance the thermal conductivity of GNPM. On the basis of phonon transmission function analysis and shifting of the first transmission function minima, it is found that the abnormal enhancement in thermal conductivity originates from the breaking of the hybridization between the resonant modes and the propagating modes by isotope doping. Our findings here provide a new means for engineering phonon transport in GNR and other nanoribbon based structures.

## Acknowledgments

N.Y. is sponsored by National Natural Science Foundation of China (No. 51576076 and No. 51711540031), Hubei Provincial Natural Science Foundation of China (2017CFA046) and Fundamental Research Funds for the Central Universities (2016YXZD006). The authors thank the National Supercomputing Center in Tianjin (NSCC-TJ) and China Scientific Computing Grid (ScGrid) for providing assistance in computations.

**Competing financial interests**: The authors declare no competing financial interests.

**Table 1.** MD simulation details and parameters.

| Method | Non- Equilibrium MD (Direct method) | | | | | |
|---|---|---|---|---|---|---|
| **Potential (TERSOFF)** | | | | | | |
| Function | $E = \frac{1}{2} \sum_i \sum_{j(\neq i)} V_{ij}$ $V_{ij} = f_C(r_{ij})[f_R(r_{ij}) + b_{ij} f_A(r_{ij})]$ $f_C(r) = \begin{cases} 1: r < R - D \\ \frac{1}{2} - \frac{1}{2}\sin(\frac{\pi}{2}\frac{r-D}{D}) : R - D < r < R + D \\ 0: \quad r > R + D \end{cases}$ $f_R(r) = A\exp(-\lambda_1 r)$ $f_A(r) = -B\exp(-\lambda_2 r) \quad\quad b_{ij} = (1 + \beta^n \varsigma_{ij}^n)^{-\frac{1}{2n}}$ $\varsigma_{ij} = \sum_{k \neq i,j} f_C(r_{ik}) g(\theta_{ijk}) \exp[\lambda_3^m (r_{ij} - r_{ik})^m]$ $g(\theta) = \gamma_{ijk}(1 + \frac{c^2}{d^2} - \frac{c^2}{[d^2 + (\cos\theta - \cos\theta_0)^2]})$ | | | | | | |
| Parameters | $m$ | $\gamma$ | $\lambda_3$ | $c$ | $d$ | $\cos\theta_0$ |
|  | 3.0 | 1.0 | 3.8e4 | 4.3484 | -0.930 | 0.72751 |
|  | $n$ | $\beta$ | $\lambda_2$ | $B$ | $R$ | $\lambda_1$ | $A$ |
|  | 1.57e-7 | 2.2119 | 430.0 | 1.95 | 0.15 | 3.4879 | 1393.6 |
| **Simulation process** | | | | | | |

| Ensemble | Setting | | Purpose |
|---|---|---|---|
| NPT | Runtime (ns) | 1ns | Relax structure |
| | Temperature (K) | 300 | |
| | Boundary condition | longitudinal, transverse, periodic, periodic | |
| NVE | Runtime (ns) | 2.5ns | |
| | Boundary condition | longitudinal, transverse, fixed, free | |
| | Thermostat | Heat source 310K | |
| | | Heat sink 290 K | |
| NVE | Runtime (ns) | 10ns | Record information |
| | Boundary condition | longitudinal, transverse, fixed, free | |
| | Thermostat | Heat source 310K | |
| | | Heat sink 290 K | |
| **Recorded physical quantity** | | | |
| Temperature | $<E> = \sum_{i=1}^{N} \frac{1}{2} m v_i^2 = \frac{1}{2} N k_B T_{MD}$ | | |
| Heat flux | $J = \frac{1}{N_t} \sum_{i=1}^{N_t} \frac{\Delta \varepsilon_i}{2 \Delta t}$ | | |
| Thermal conductance | $\kappa = -\frac{J}{W \cdot d \cdot \nabla T}$ | | |